%% file: main.tex
\documentclass[journal,transmag]{IEEEtran}

\ifCLASSINFOpdf
\else
\fi

\usepackage{graphicx}
\usepackage{hyperref}
\usepackage{bm}
\usepackage{amsfonts}
\usepackage{amsmath}
\usepackage{tikz}
\usepackage{pgfplots}

\usepackage{subcaption}
\usepackage{eso-pic}

\newcommand{\ddx}[2]{%
	\frac{\mathrm{d}#1}{\mathrm{d}#2}
}

\begin{document}
\title{Gradient-Informed Machine Learning in Electromagnetics}

\author{\IEEEauthorblockN{Matteo Zorzetto\IEEEauthorrefmark{1},
Merle Backmeyer\IEEEauthorrefmark{2}, 
Michael Wiesheu\IEEEauthorrefmark{2},
\\ Riccardo Torchio\IEEEauthorrefmark{1,3}, Fabrizio Dughiero \IEEEauthorrefmark{1}, and 
Sebastian Sch\"ops\IEEEauthorrefmark{2}}
\IEEEauthorblockA{\IEEEauthorrefmark{1}Department of Industrial Engineering, University of Padova, 35131  Padova, Italy}
\IEEEauthorblockA{\IEEEauthorrefmark{2}Computational Electromagnetics Group, Technische Universit\"at Darmstadt, 64289 Darmstadt, Germany}%
\IEEEauthorblockA{\IEEEauthorrefmark{3}Department of Information Engineering, University of Padova, 35131  Padova, Italy}
\thanks{Received 23 January 2026; revised 8 April 2026 and 5 June 2026;
	accepted 6 June 2026. Corresponding author: M. Zorzetto (e-mail: matteo.
	zorzetto@phd.unipd.it).
	
	Color versions of one or more figures in this article are available at
	https://doi.org/10.1109/TMAG.2026.3702291.
	
	Digital Object Identifier 10.1109/TMAG.2026.3702291}}

\markboth{Gradient-Informed Machine Learning in Electromagnetics}%
{Zorzetto \MakeLowercase{\textit{et al.}}: Gradient-Informed Machine Learning in Electromagnetics}

\IEEEtitleabstractindextext{%
\begin{abstract}
Simulation techniques such as the finite element method are essential for designing electrical devices, but their computational cost can be prohibitive for repeated or real-time computations. Projection-based model order reduction techniques mitigate this by reducing the model size and complexity, yet face challenges when extended to nonlinear or non-affine parametric models. In this work, Isogeometric Analysis (IGA) is combined with proper orthogonal decomposition and Gaussian process regression to construct a non-intrusive surrogate model of a parametric nonlinear model of a permanent magnet synchronous machine. The differentiable nature of IGA allows for computationally efficient extraction of parametric sensitivities, which are leveraged for gradient-enhanced surrogate modeling.
\end{abstract}

}

\AddToShipoutPicture*{
	\footnotesize\sffamily\raisebox{0.8cm}{\hspace{1.4cm}\fbox{
			\parbox{\textwidth}{
				© 2026 IEEE. Personal use of this material is permitted. Permission from IEEE must be obtained for all other uses, including reprinting/republishing this material for advertising or promotional purposes, collecting new collected works for resale or redistribution to servers or lists, or reuse of any copyrighted component of this work in other works.
			}
	}}
}

\maketitle

\IEEEdisplaynontitleabstractindextext

\IEEEpeerreviewmaketitle

\section{Introduction}
\IEEEPARstart{S}{imulation} techniques offer powerful tools for analyzing, designing, and monitoring electrical devices. Methods such as the finite element method convert physical equations into large systems of algebraic equations, which can then be solved using numerical linear algebra. However, the computational cost associated with solving these systems can become a limiting factor when repeated computations are required, such as in system-level simulations, design optimization, or real-time control.
Projection-based Model Order Reduction (MOR) techniques address this problem by projecting the system matrices onto a lower-dimensional subspace, obtaining a computationally efficient Reduced-Order Model (ROM). Approaches such as balanced truncation, moment matching, and Proper Orthogonal Decomposition (POD)~\cite{benner_PMOR_survey, chatterjeeIntroductionProperOrthogonal2000} are particularly effective for linear systems and differ by how they find the reduced basis. However, their extension to nonlinear and parametric models is challenging and intrusive~\cite{benner_PMOR_survey}. Algorithms such as the discrete empirical interpolation method can treat nonlinear terms~\cite{DEIM,henneronModelOrderReductionMultipleInput2015,zorzettoReducedOrderModeling2024}, but their application requires access to the nonlinear representation of the system, which is not guaranteed, especially when commercial software is involved~\cite{ortaliGaussianProcessRegression2021}. Moreover, although projection-based methods can be extended to handle affine parametric dependencies, producing a parametric ROM that preserves the structure of the original model, certain types of dependencies remain difficult to accommodate~\cite{benner_PMOR_survey}. In particular, geometric parameter variations often violate the assumptions required for an affine decomposition, making them poorly suited for standard projection-based MOR frameworks.
In recent years, machine learning techniques have been increasingly employed to overcome these challenges by learning the relationship between system parameters and quantities of interest directly from data. In this framework, the high-dimensional field variables can be decomposed using methods such as POD~\cite{henneronSurrogateModelBased2020,zorzettoProperOrthogonalDecomposition2025} or autoencoders~\cite{zorzettoPodMultiphysics, pichiGraphConvolutionalAutoencoder2024a}, resulting in a compact set of latent variables that capture the dominant dynamics of the system. Machine learning models such as neural networks~\cite{backmeyerLearningElectromagneticFields2025}, radial basis function interpolation~\cite{henneronSurrogateModelBased2020}, and Gaussian process regression (GPR)~\cite{ortaliGaussianProcessRegression2021,zorzettoPodMultiphysics,zorzettoProperOrthogonalDecomposition2025}, can then be trained to map the system parameters to this reduced space, enabling the reconstruction of the full-order field with significantly lower computational cost. In this work, building upon~\cite{zimmermannGradientenhancedSurrogateModeling2013}, we propose a Gradient-Enhanced (GE) surrogate modeling framework that combines Isogeometric Analysis (IGA) with POD and GPR (aka Kriging), extending the previously proposed Gradient-Free (GF) approach~\cite{backmeyerLearningElectromagneticFields2025}. Leveraging the exact geometric parametrization of IGA, the proposed approach enables the non-intrusive and computationally efficient extraction of the state vector sensitivities with respect to geometric parameters, quantities not readily accessible in finite element method formulations. This gradient information can be incorporated into the reduced basis to construct a more expressive subspace for representing the state vector. Moreover, it can be used within gradient-enhanced interpolation schemes, such as GPR, to improve the predictive accuracy of the surrogate in the latent space with respect to gradient-free approaches~\cite{zimmermannGradientenhancedSurrogateModeling2013}.
\if
In this context, Isogeometric Analysis (IGA) offers a significant advantage over the standard finite element method, since we can keep the spatial mapping consistent to enable calculation of gradients. This allows for computationally efficient extraction of the state vector sensitivities with respect to geometric parameters. This derivative information can be leveraged to enhance the surrogate model. It can be incorporated into the reduced basis to construct a more expressive subspace for representing the state vector. Moreover, it can be used within gradient-enhanced interpolation schemes, such as GPR (aka Kriging), to improve the predictive accuracy of the surrogate in the latent space with respect to gradient-free approaches~\cite{zimmermannGradientenhancedSurrogateModeling2013}.
In this work, the Gradient-Free (GF) IGA approach, previously combined with POD interpolation in~\cite{backmeyerLearningElectromagneticFields2025}, is further enhanced by incorporating the Gradient-Enhanced (GE) methodology introduced in~\cite{zimmermannGradientenhancedSurrogateModeling2013}. 
\fi
The resulting method is capable of generating accurate and efficient surrogate models of electrical devices with nonlinear material behavior and geometric parameter variations in a non-intrusive manner. The effectiveness of the approach is demonstrated on a Permanent-Magnet Synchronous Machine (PMSM) through two studies. First, POD is combined with gradient-enhanced GPR to construct a surrogate model of the electromagnetic field distribution. Moreover, the derivative information is directly leveraged to build surrogate models for Key Performance Indicators (KPIs) such as torque.

\section{Motor Model}\label{sec:motor}
The method is applied to a PMSM from~\cite{Pahner_1998ab, Komann_2024aa}, shown in Figure~\ref{fig:motor_geometry}, function of the geometric design variables $\mathbf{p} = [\operatorname{MH}, \operatorname{MW}, \operatorname{MAG}, \operatorname{\theta_1}]$, and solved for a fixed number of rotation angles $\alpha$. The iron cores are shown in gray, the (homogenized) copper slots in red, the rotor magnet in green, and the air gap and air pockets in blue. The physical model is described in detail in~\cite{Komann_2024aa}, while the parameterization is discussed in~\cite{backmeyerLearningElectromagneticFields2025}. For brevity, only the key aspects are outlined here.
\begin{figure}[t]
    \centering
    \scalebox{0.4}{
    \input{figures/motor/PMSMgeometry.tikz}
    }
	\caption{Parametrization of the PMSM geometry, based on~\cite{Komann_2024aa}.}
	\label{fig:motor_geometry}
\end{figure}
\subsection{Governing Equations}
 For the laminated PMSM, %
 the parameter-dependent magnetic vector potential in the rotor and stator, $A_{z,\mathrm{R}}$ and $A_{z,\mathrm{S}}$ respectively, is obtained from the two-dimensional magnetostatic Poisson problem~\cite{Salon_1995aa, Wiesheu_2024aa}
\begin{align}
        \label{eq:magnetostatics_2D_1}
        \nabla \cdot (\nu \nabla A_{z,\mathrm{R}}^{(\mathbf{p})}) &= \nu \nabla \cdot {\mathbf{B}_\text{rem}^\perp}^{(\mathbf{p})}
        \quad&& \text{in } \Omega_{\mathrm{R}}^{(\mathbf{p})},
        \\
        \label{eq:magnetostatics_2D_2}
        \nabla \cdot (\nu \nabla A_{z,\mathrm{S}}^{(\mathbf{p})}) &= -J_{z,\text{src}} && \text{in } \Omega_{\mathrm{S}}^{(\mathbf{p})},
\end{align}
driven by the source current density  $J_{z,\text{src}}$. The 2D remanent flux density of the permanent magnets ${\mathbf{B}_\text{rem}^\perp}^{(\mathbf{p})} = B_\text{rem}\left[-\sin(\alpha),\cos(\alpha)\right]^{\top}$ is computed from the remanence magnitude $B_\text{rem}$ and the rotation angle $\alpha$. It may also depend on the parameter vector. The reluctivity $\nu$ is set linear for air, copper, and the magnet. For the iron parts, we have $\nu=\nu(B)$ given through a nonlinear $\mathrm{B\!-\!H}$ curve. The magnetic flux density is given by $\mathbf{B}^{(\mathbf{p})} = \nabla \times \left(A_z^{(\mathbf{p})} \mathbf{e}_z\right)$. Problem (\ref{eq:magnetostatics_2D_1}--\ref {eq:magnetostatics_2D_2}) is complemented with standard boundary conditions, and the coupling across $\Gamma_\text{ag}=\bar\Omega_{\mathrm{R}}^{(\mathbf{p})}\cap\bar\Omega_{\mathrm{S}}^{(\mathbf{p})}$ is enforced by a Lagrange multiplier and additional coupling conditions (see~\cite{Egger_2022ab}). The electromagnetic torque is computed from the field solution; thus it also depends on the parametric realization of the problem~\cite{Salon_1995aa}. 
\subsection{IGA Discretization}

The discretization of the geometry as well as the solution is performed using Non-Uniform Rational B-Spline basis functions, following the principles of IGA~\cite{Buffa_2010aa}. The physical domain is described by a multi-patch parametrization, decomposing it into a collection of subdomains, each with a corresponding projection map to the same reference domain~\cite{Buffa_2015aa}. Despite different geometrical realizations, the reference domain stays the same, which is essential if the parametric problem is learned in terms of its coefficients. For reasonable parametric changes, no remeshing is required, and the learned coefficients belong to the same nodes in the reference domain which are then mapped to the corresponding physical domain~\cite{backmeyerLearningElectromagneticFields2025}.
Problem (\ref{eq:magnetostatics_2D_1}--\ref {eq:magnetostatics_2D_2}) is discretized with B-splines following the standard Ritz-Galerkin approach. The magnetic vector potential in the reference domain is approximated by
\begin{equation}
A_z^{(\mathbf{p})}(\textbf{x})
\approx 
\sum\nolimits_{i=1}^N B_i^q(\mathbf{p},\textbf{x})\; u_i
\label{eq:AnsatzTestFunctions}
\end{equation}
where  $u_i$ are the unknown coefficients and $B_i^q(\mathbf{p},\textbf{x})$ are the spline basis functions of degree $q$. %
The physical problem results in a nonlinear system of equations dependent on the geometrical parameters~$\mathbf {p}$:
\begin{equation}
    \mathbf{A}\left(\mathbf{p},\mathbf{u}\right)\mathbf{u} = \mathbf{b},
    \label{eq:system}
\end{equation}
where $\mathbf{u}$ is the state vector, including variables for the rotor and stator, as well as Lagrange multipliers for rotor-stator coupling. $\mathbf{A}\left(\mathbf{p},\mathbf{u}\right)$ is assembled with stiffness matrices and coupling terms that depend on geometrical parameters and the state vector due to the presence of nonlinear materials. The vector~$\mathbf{b}$ describes the excitations of rotor and stator that are given by the permanent magnet and the coils. 
For more details on constructing the matrix system \eqref{eq:system} the reader is referred to~\cite{Wiesheu_2024aa}.
As outlined in~\cite{Wiesheu_2024aa}, analytical computation of the gradients is feasible through the adjoint method.
Our implementation of IGA~\cite{Backmeyer_2025ad} not only provides the solution vector $\mathbf{u}\left(\mathbf{p_*}\right)$ but also enables efficient evaluation of its sensitivities with respect to each component of $\mathbf{p}$, as briefly discussed in the following.

\subsection{Calculation of Derivatives}

The sensitivities with respect to the parameters are obtained by applying the adjoint method. After solving \eqref{eq:system} iteratively with the Newton-Raphson scheme, the total derivative of the torque $T$ is obtained by 
\begin{equation}
    \ddx{T}{\mathbf{p}} = \mathbf{z}^\top \left(\frac{\partial (\mathbf{A}\mathbf{u})}{\partial\mathbf{p}} -\frac{\partial\mathbf{b}}{\partial\mathbf{p}}\right)
    \label{eq:gradient}
\end{equation}
introducing the adjoint variable $\mathbf{z}$ with
\begin{equation}
    \mathbf{J}^\top\mathbf{z} = -\frac{\partial T}{\partial\mathbf{u}},
    \label{eq:adjoint}
\end{equation}
where $\mathbf{J}$ is the Jacobian matrix of \eqref{eq:system} after the Newton solve. This makes the evaluation of gradients very efficient, as even if the system is nonlinear, the adjoint approach just requires one linear system solve to obtain the gradient. 

This leaves only the calculation of partial derivatives needed to evaluate \eqref{eq:gradient} and \eqref{eq:adjoint}. The partial derivative with respect to the torque needed in \eqref{eq:adjoint} is simple, because the torque is a closed-form expression directly depending on $\mathbf{u}$, see \cite[Eq.~30]{Wiesheu_2024aa}.

The partial derivatives in \eqref{eq:gradient} are more involved, as the changes in system stiffness matrix and right-hand side vector with respect to parameters need to be derived. For standard Finite Elements, this might prove difficult, since a new geometry usually involves remeshing, making it impossible to track the contributions in the system matrix and vector. This is different in the IGA case, where the geometry mapping is predefined by patches and their corresponding control points. There are efficient analytical expressions for calculating the derivatives of $\mathbf{A}$ and $\mathbf{b}$ with respect to the control points~\cite{Wiesheu_2024aa}. The derivatives with respect to the parameters are then obtained in a postprocessing step by realizing that changing one parameter directly corresponds to changing all control points that are related to the parameter. Mathematically, this corresponds to applying the chain rule. This makes the evaluation of sensitivities computationally inexpensive, even for a large parameter set. If an additional parameter should be optimized, it only needs to be determined how the control points depend on it.

\section{Surrogate Modeling}
The surrogate modeling approach employed in this work consists of two main components: POD, used to obtain a low-dimensional representation of the solution fields, and GPR, used to model the dependence of the reduced coefficients on the parameters. Following~\cite{zimmermannGradientenhancedSurrogateModeling2013}, both POD and GPR are enriched with gradient information from the IGA model to improve accuracy. 

\subsection{Proper Orthogonal Decomposition}\label{sec:pod}
POD finds a reduced representation of the solution starting from the snapshot matrix, formed by concatenating a set of solutions of the problem:
\begin{equation}\label{eq:snapmat}
    \mathbf{U} = \left[\bm{u}(\mathbf{p}_1), \ldots, \bm{u}(\mathbf{p}_{n_s})\right] \in \mathbb{R}^{n\times n_s},
\end{equation}
where $n$ is the number of degrees of freedom, while $n_s$ is the number of samples in the set. The main assumption behind POD is that the solution vector can be approximated by a linear combination of $r \ll n$ basis vectors~$\bm{\phi}$, weighted by parameter-dependent coefficients $\bm{\xi}(\mathbf{p})$. The reduced basis $\bm{Q} = \left[\bm{q}_1, \cdots, \bm{q}_r \right] \in \mathbb{R}^{n\times r}$ is commonly obtained by taking the first $r$ left singular vectors from the Singular Value Decomposition (SVD) of $\mathbf{U}$. Following~\cite{backmeyerLearningElectromagneticFields2025}, POD is performed with a weighted inner product obtained via a weighting matrix $\bar{\mathbf{A}} = \mathbf{A}(\bar{\mathbf{p}}, \mathbf{0})$, chosen as the system matrix from \autoref{eq:system}, evaluated with homogeneous unit material at the expected parameter combination $\bar{\mathbf{p}}$. Although $\mathbf{A}$ depends on the parameters, it is kept constant to simplify the basis computation. With this weighted inner product, the POD modes are obtained by solving the generalized eigenvalue problem
\begin{equation}\label{eq:eig}
    \mathbf{U}^\top \bar{\mathbf{A}} \mathbf{U} \, \bm{\phi}_i = \lambda_i \, \bm{\phi}_i, 
    \qquad i = 1, \ldots, r,
\end{equation}
and defining the reduced basis vectors as
\begin{equation}
    \bm{q}_i = \frac{1}{\sqrt{\lambda_i}} \, \mathbf{U} \bm{\phi}_i, 
    \qquad i = 1, \ldots, r.
\end{equation}
The reduced coefficients can then be computed for a generic parameter combination $\mathbf{p}^*$ by

\begin{equation}\label{eq:sol_to_coeff}
\bm{\xi}(\mathbf{p}
     ^*) = \mathbf{Q}^\top\bar{\mathbf{A}}\mathbf{u}(\mathbf{p}^*).
\end{equation}
Truncating the basis at a fixed value $r$ introduces a reconstruction error, which can be monitored by:

\begin{equation}\label{eq:erel}
    \epsilon_\text{rel}^2(\mathbf{p}^*) = \dfrac{\left(\mathbf{u}(\mathbf{p}^*)-\mathbf{Q}\tilde{\bm{\xi}}(\mathbf{p}^*)\right)^\top\mathbf{A}\left(\mathbf{u}(\mathbf{p}^*)-\mathbf{Q}\tilde{\bm{\xi}}(\mathbf{p}^*)\right)}{\mathbf{u}(\mathbf{p}^*)^\top\mathbf{A}\mathbf{u}(\mathbf{p}^*)}.
\end{equation}

After dimensionality reduction, machine learning approaches such as GPR can be used to find a function $\mathbf{f}$, mapping $\mathbf{p}$ to $\bm{\xi}$, from which the field distribution can be reconstructed by:
\begin{align}\label{eq:coeff_to_sol}
\mathbf{u}(\mathbf{p}
     ^*) \approx \tilde{\mathbf{u}}(\mathbf{p}
     ^*) =  \mathbf{Q} \bm{\xi}, && \bm{\xi} \approx \mathbf{f}(\mathbf{p}^*).
\end{align}
The accuracy of the POD reconstruction is directly influenced by the size of the basis. While increasing $r$ improves accuracy, it also raises computational costs, requiring a compromise between efficiency and error. A typical approach is to choose $r$ so that the relative cumulative energy:
\begin{equation}
    \dfrac{\sum_{k=1}^r \lambda_k}{\sum_{k=1}^{n_s} \lambda_k} > \text{tol},
\end{equation}
is greater than a user-defined tolerance, typically set to be greater than 90\% \cite{backmeyerLearningElectromagneticFields2025}\cite{Willcox_projection_based_MOR}. In this work, the rotation angle is treated explicitly by solving each rotor geometry for a set of discrete rotation angles. To achieve faster inference, rather than introducing it as an additional parameter, POD is applied to an expanded snapshot matrix obtained by concatenating the solutions corresponding to the discrete rotor positions. The resulting snapshot vectors therefore have dimension $n_\alpha n$. The proposed treatment of the rotor position does not increase the size of the eigenvalue problem \eqref{eq:eig}, since its dimension is determined by the number of sampled parameter combinations $n_s$ rather than by the size of the snapshot vectors. Furthermore, it does not affect the complexity of the subsequent GPR fit. Since the regression model is trained only on the reduced coefficients $\bm{\xi}$, its computational cost is independent of the number of discrete rotor positions $n_\alpha$.

When gradient information is available, a more general POD basis can be derived by augmenting the snapshot matrix $\mathbf{U}$~\cite{zimmermannGradientenhancedSurrogateModeling2013}:
\begin{equation}\label{eq:snapmat_aug}
    \mathbf{U}_\text{aug} = \left[\bm{u}(\mathbf{p}_1),\bm{u}_1(\mathbf{p}_1),\ldots, \bm{u}_{n_p}(\mathbf{p}_1), \ldots, \bm{u}_{n_p}(\mathbf{p}_{n_s})\right],
\end{equation}
where $n_p$ is the number of parameters and $\bm{u}_i(\mathbf{p}_j)$ stands for~\cite{zimmermannGradientenhancedSurrogateModeling2013}:
\begin{equation}
    \bm{u}_i(\mathbf{p}_j) = \bm{u}(\mathbf{p}_j)+ h\cdot\dfrac{\partial \bm{u}}{\partial p_{i}}(\mathbf{p}_j).
\end{equation}
where $h$ is a small value, chosen here to be $10^{-3}$.
Since the magnetic vector potential is learned as coefficients with respect to spline basis functions, the solution inherits the properties of the basis. Therefore, the proposed approach preserves several relevant structural and physical properties by construction. For instance, the
resulting magnetic flux densities are solenoidal and exhibit the appropriate normal continuity across material interfaces, while the magnetic vector potentials and electric fields exhibit the corresponding
tangential continuity. Moreover, all field energies are non-negative by construction. In general, however, the resulting fields do not satisfy the governing equations exactly, neither in strong nor in weak form, and other derived quantities may lack certain physical properties. This distinguishes the proposed non-intrusive POD-regression framework from many classical projection-based MOR techniques \cite{Volkwein_2013aa,Henneron_2014aa}, which may preserve additional equation-based properties but are typically intrusive and require the solution of nonlinear systems for each evaluation in parameter space.

\subsection{Gaussian Process Regression}
This section briefly outlines GPR and its extension to the gradient-enhanced case. A more detailed presentation can be found in~\cite{GPR_tutorial}.
A Gaussian process $\mathcal{GP}$ is a distribution over functions and is defined by a mean function $\mu(\mathbf{x})$ and a covariance function $k(\mathbf{x},\mathbf{x}')$~\cite{GPR_tutorial} as
\begin{equation}
	f(\mathbf{x}) \sim \mathcal{GP}\left(\mu(\mathbf{x}), k(\mathbf{x},\mathbf{x}')\right),
\end{equation}
where the mean function 
\begin{align}
    \mu(\mathbf{x}) &= \mathbb{E}[f(\mathbf{x})]
\end{align}
represents the expected function value for input $\mathbf{x}$, while the covariance function 
\begin{align}
    k(\mathbf{x}, \mathbf{x}') &= \mathbb{E}\left[\left(f(\mathbf{x})-\mu(\mathbf{x})\right)\left(f(\mathbf{x}')-\mu(\mathbf{x}')\right)\right]
\end{align}
models the relationship between function values at different input points $\mathbf{x}$ and $\mathbf{x}'$~\cite{GPR_tutorial, santnerDesignAnalysisComputer2018}.
A set of training data $(\bm{X}_t, \mathbf{y}_t)$ is considered, where $\bm{X}_t$ corresponds to the parameter vectors $\mathbf{p}$ and $\mathbf{y}_t$ to the associated reduced coefficients $\bm{\xi}$. A kernel is then selected, such as the radial basis function:
\begin{equation}
	k(\mathbf{x}, \mathbf{x}') = \sigma_f^2 \exp\left( -\dfrac{||\mathbf{x} - \mathbf{x}'||^2}{2\lambda^2} \right),
\end{equation}
where $\sigma_f$ is the signal variance and $\lambda$ is the length scale. Assuming zero mean, the function value for an unseen input $\mathbf{x}_*$ can then be obtained by computing the posterior mean of the distribution:
\begin{equation}
	\mu(\mathbf{x}_*) = \mathbf{K}(\mathbf{x}_*,\bm{X}_t)\left[\mathbf{K}(\bm{X}_t,\bm{X}_t)+\sigma_\epsilon^2\bm{I}\right]^{-1}\mathbf{y}_t,
\end{equation}
where $\sigma_\epsilon^2$ is the assumed noise level of the observations, $\mathbf{I}$ is the identity matrix and entries in the covariance matrix $\mathbf{K}$ are defined by evaluating the kernel function
\begin{equation}
	\mathbf{K}(\bm{X}_a,\bm{X}_b) = \left\{k(\mathbf{x}^a_i,\mathbf{x}^b_j)\right\}_{ij},
\end{equation} 
where $\bm{X}_a$ and $\bm{X}_b$ represent arbitrary sets of input points. One of the main advantages of GPR compared to other interpolation approaches is the availability of a confidence interval, which can be estimated by the variance function
\begin{equation}
	\begin{aligned}
		\sigma^2(\mathbf{x}_*) &= k(\mathbf{x}_*, \mathbf{x}_*) \\
		&\quad - \mathbf{K}(\mathbf{x}_*, \mathbf{X}_t) \left[\mathbf{K}(\mathbf{X}_t, \mathbf{X}_t) + \sigma_\epsilon^2 \mathbf{I}\right]^{-1} \mathbf{K}(\mathbf{X}_t, \mathbf{x}_*).
	\end{aligned}
\end{equation}
The extension of GPR to the gradient-enhanced case (GEGPR) is performed by modeling the solution and its partial derivatives jointly. For each training point $\mathbf{x}$, the solution is extended from $y = f(\mathbf{x})$ to: 

\begin{equation}
    \mathbf{y} = \left[f(\mathbf{x}),\dfrac{\partial f}{\partial \mathrm{x}_i}(\mathbf{x}) , \ldots , \dfrac{\partial f}{\partial \mathrm{x}_{n_p}}(\mathbf{x})\right].
\end{equation}

\begin{figure}[t]
    \centering
    \begin{subfigure}[b]{0.49\linewidth}
    \centering
    \includegraphics[width=\textwidth]{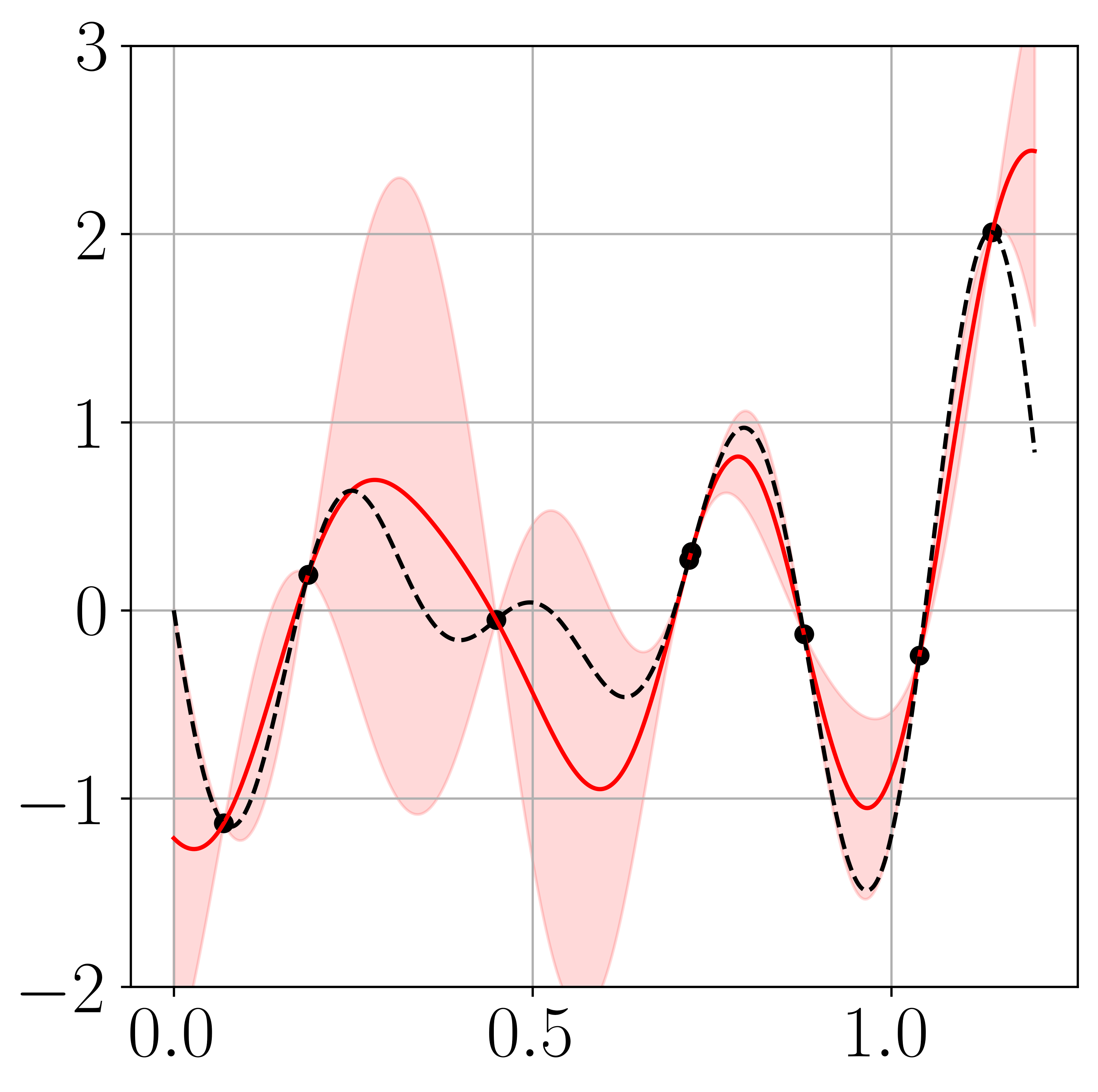}
    \caption{Gradient-free}
    \end{subfigure}
    \hfill
    \begin{subfigure}[b]{0.49\linewidth}
    \centering
    \includegraphics[width=\textwidth]{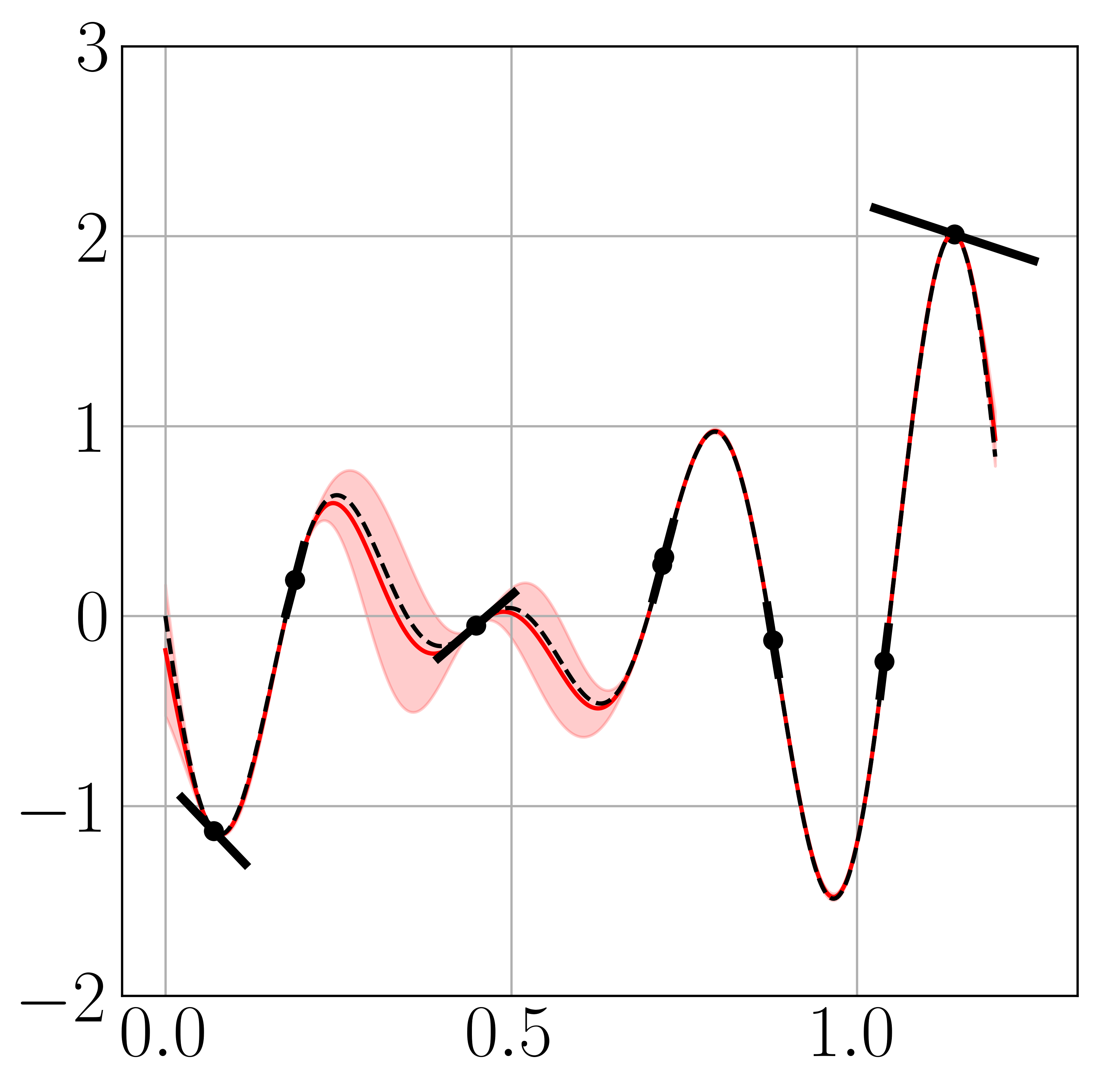}
    \caption{Gradient-enhanced}
    \end{subfigure}
    \caption{Comparison between GFGPR and GEGPR. Black markers indicate the samples of the true function (represented by the dashed black line) used to fit the models. Red lines correspond to the prediction, while the shaded area is the 95\% confidence interval.}
    \label{fig:GPR_withd_example}
\end{figure}

Matrix $\mathbf{K}$ is then modified to include entries corresponding to the covariance between function values $k(\mathbf{x},\mathbf{x}')$, covariance between function values and each partial derivative:

\begin{equation}
    \dfrac{\partial k(\mathbf{x},\mathbf{x}')}{\partial \mathrm{x}_i},
\end{equation}
and covariance between partial derivatives~\cite{rasmussenGaussianProcessesMachine2008a}: 

\begin{equation}
\dfrac{\partial^2 k(\mathbf{x},\mathbf{x}')}{\partial \mathrm{x}_i\partial \mathrm{x}_j'}.
\end{equation}
The primary computational bottleneck of GPR is the inversion of the covariance matrix, which scales asymptotically as $\mathcal{O}\left(n_s^3\right)$. In GEGPR, the covariance matrix is expanded to incorporate both function values and their gradients, resulting in a size of $n_s(1 + n_p)$. Consequently, the computational cost increases to $\mathcal{O}\left(n_s^3 (1+n_p)^3\right)$~\cite{boneGradientenhancedDeepGaussian2025}. Although this higher scaling may initially appear disadvantageous, the inclusion of gradient information effectively augments the dataset, increasing the number of observations from $n_s$ to $n_s(1 + n_p)$. The main advantage of combining GEGPR with IGA is that sensitivity information can be obtained at a significantly lower cost, i.e., one linear system for all derivatives, than generating additional high-fidelity samples, requiring multiple linear systems to be solved. As a result, gradient information can enrich the training dataset with only a modest increase in the data-generation cost. Nevertheless, for large datasets and high-dimensional parameter spaces, the computational complexity of GPR can become prohibitive, in which case sparse approximations may be employed to improve scalability~\cite{quinonero2005unifying}.
In \autoref{fig:GPR_withd_example}, the standard GPR is compared to the gradient-enhanced version for modeling a simple one-dimensional function. The inclusion of derivative information improves the function estimate and tightens the confidence bounds.

\section{Results}
A parametrized IGA model of the geometry in \autoref{fig:motor_geometry} was developed, discretizing the physical equations into a nonlinear system of 6177 variables, which is a function of the geometrical parameters presented in \autoref{tab:bounds}. To monitor the accuracy of the approach in data-scarce conditions, Sobol sampling was employed to generate training datasets of increasing size. Four nested sets of 16, 32, 64, and 128 samples were initially generated within the bounds in \autoref{tab:bounds}. All combinations leading to a magnet exceeding the geometric bounds were excluded from the dataset. Additionally, combinations in which the angle between the magnet edge and the adjacent rotor iron boundary was less than $10^\circ$ were removed to eliminate geometries where the flux barrier becomes excessively skewed. All geometries were solved for a set of $n_\alpha = 14$ equidistant rotation angles in [0, 20]$^\circ$. The model was simulated on a workstation equipped with a 3.10 GHz Intel\textsuperscript{®} Xeon\textsuperscript{®} processor, with an average time of 9 seconds per solve, resulting in approximately 120 seconds per data point due to the 14 rotation angles. The resulting training datasets are summarized in \autoref{tab:times}, while all models were tested on the same collection of 30 unseen combinations. The torque was subsequently obtained through postprocessing of the solution, yielding a distribution with a mean of 0.86 Nm and a standard deviation of 0.35 Nm.

\begin{table}[t]
\caption{Parameter bounds}\label{tab:bounds}
\centering
\begin{tabular}{llcc}
Name & unit & lower bound & upper bound \\ \hline
MH             & mm   & 2   & 12  \\
MW             & mm   & 8   & 22  \\
MAG            & mm   & 5   & 15  \\
$\theta_1$     & °    & 15  & 23 \\ \hline
\end{tabular}
\end{table}

\begin{table}[]
\centering
\caption{Times for dataset generation and model fitting for different dataset partitions. Each sample is associated with 14 simulations with increasing rotation angle.}\label{tab:times}
\begin{tabular}{lllll}
$n_s$ & 15   & 31   & 61  & 119  \\
\hline
Simulation time (min) & 30  & 62  & 123 & 239 \\
Field GFGPR  (s)  & 8.5 & 8.7  & 7.8 & 11.2 \\
Field GEGPR  (s)  & 16.8 & 25.2 & 53.4  & 124.5 \\
Torque GFGPR  (s) & 13.2  & 9.6 & 10.9 & 13.3  \\
Torque GEGPR (s) & 23.7  & 55.0  & 60.7 & 103.0 \\ \hline
\end{tabular}
\end{table}

\subsection{Field Decomposition}
\begin{figure}[!t]
        \centering
        \includegraphics[width=\linewidth]{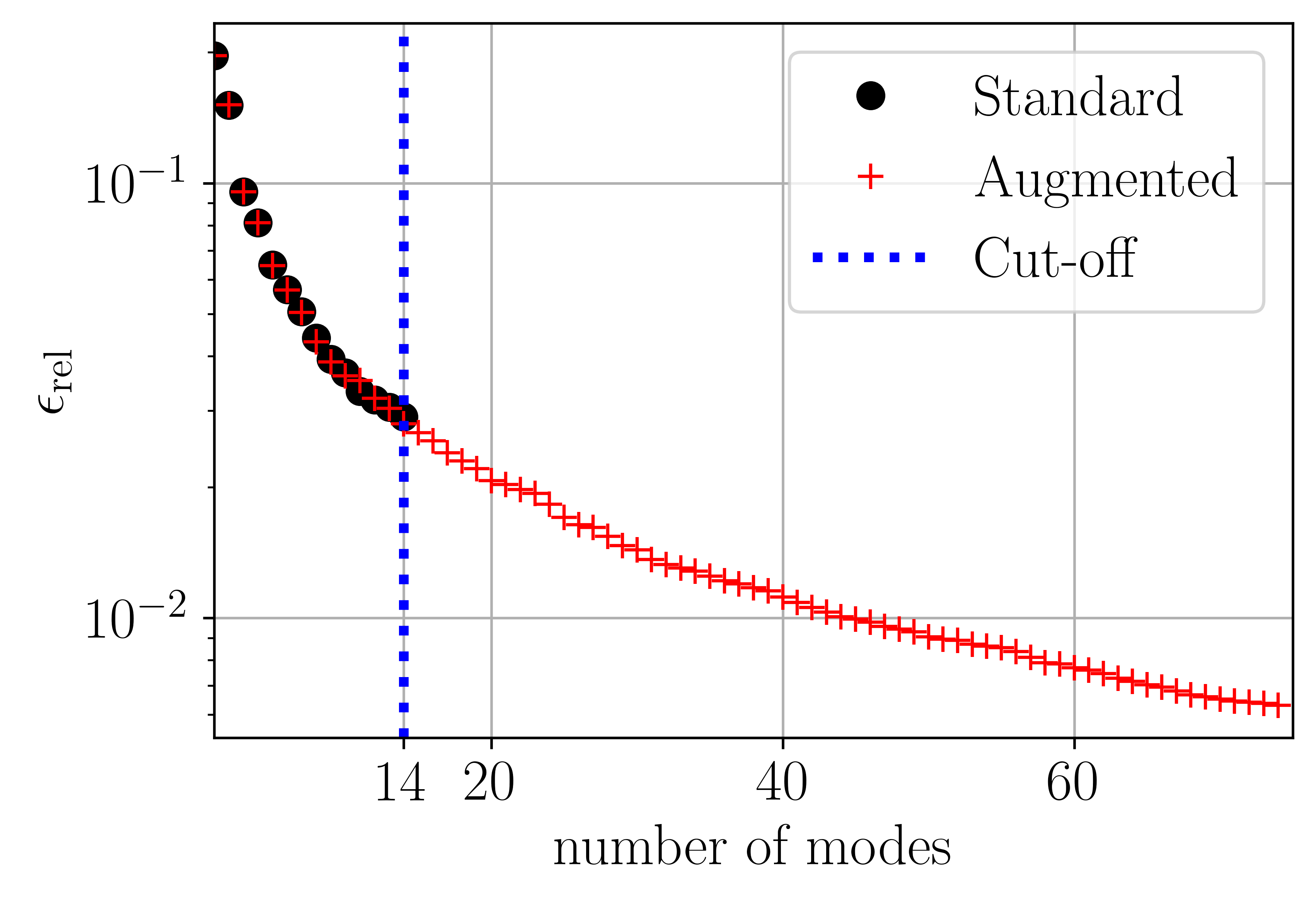}
        \caption{Average relative $L_2$ reconstruction error of the magnetic flux density for the gradient-free and gradient-enhanced basis for $n_s$ = 15.}
        \label{fig:err_vs_size_small}
\end{figure}

\begin{figure}[t]
        \centering
        \includegraphics[width=\linewidth]{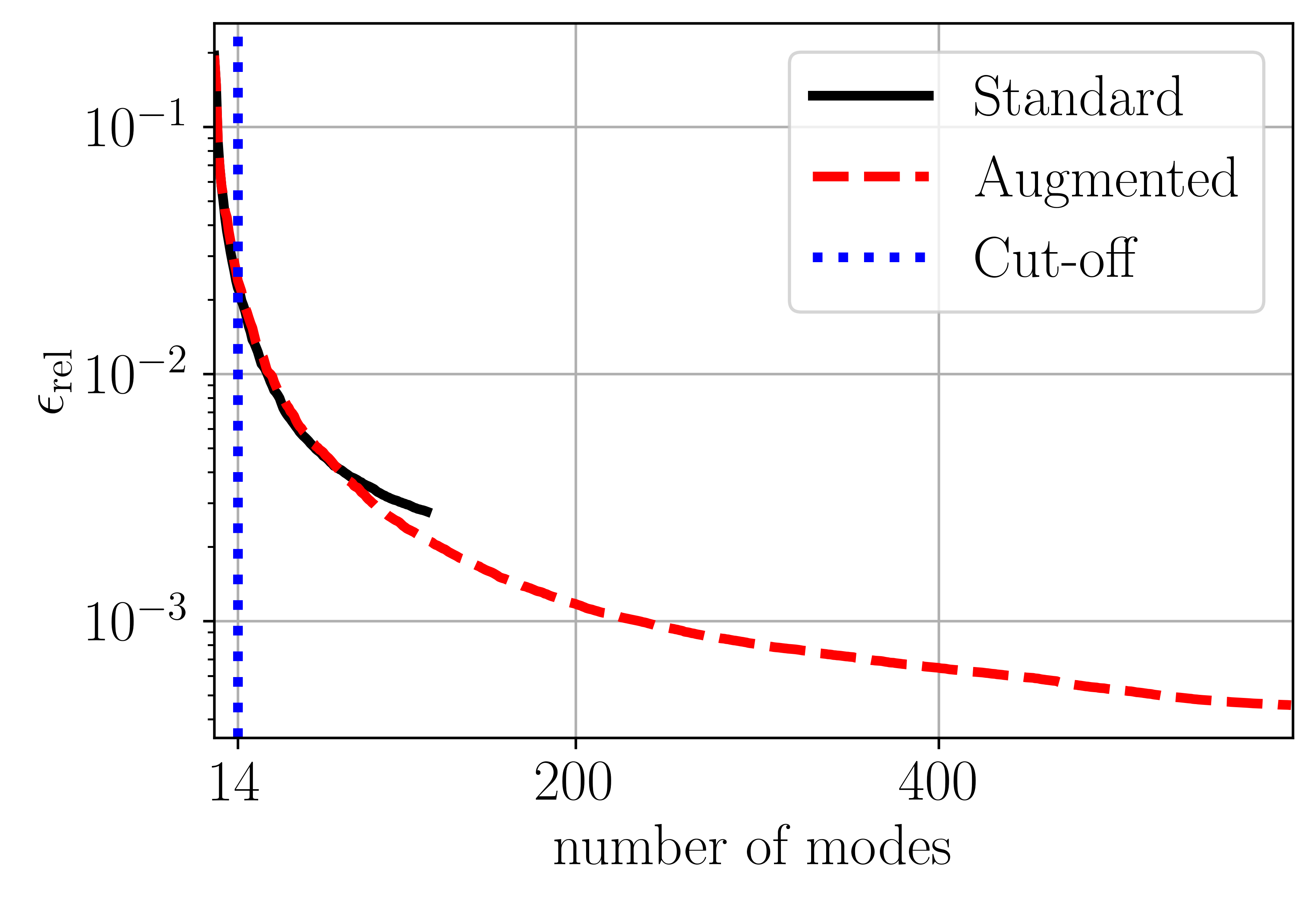}
        \caption{Average relative $L_2$ reconstruction error of the magnetic flux density for the gradient-free and gradient-enhanced basis for $n_s$ = 119.}
        \label{fig:err_vs_size_big}
\end{figure}

For each dataset partition, the standard and augmented snapshot matrices were realized according to \eqref{eq:snapmat} and \eqref{eq:snapmat_aug}. To compare the two approaches, their accuracy was evaluated by monitoring the average relative $L_2$ error of the magnetic flux density through \eqref{eq:erel} over the test dataset, computed using $\bar{\mathbf{A}}$.
In \autoref{fig:err_vs_size_small}, the relative error is presented for the POD basis trained on the smallest dataset of 15 samples. In this scenario, the augmented snapshot matrix allows for an increased basis size without requiring additional physical samples, thus leading to a smaller reconstruction error. It should be noted, though, that a rank of 14 constitutes an edge case where the training data is perfectly recovered, while generalization to new samples is not guaranteed. In contrast, as presented in \autoref{fig:err_vs_size_big}, when the number of available samples grows, an improvement in the reconstruction error begins to appear for truncations close to $n_s$. This behavior is attributed to information redundancy in the non-augmented dataset, which limits generalization when $r$ approaches $n_s$. Since the goal of MOR is to obtain accurate but compact models, large truncation levels are undesirable. Consequently, for this application, snapshot augmentation provides only marginal benefits, limited to the smallest dataset. Therefore, the standard POD basis is adopted in all scenarios, fixing the number of modes to 14, ensuring a fair comparison for the following latent-space modeling.

\subsection{Field reconstruction}\label{sec:field_rec}
\begin{figure}[!t]
	\centering
	\includegraphics[width=0.9\linewidth]{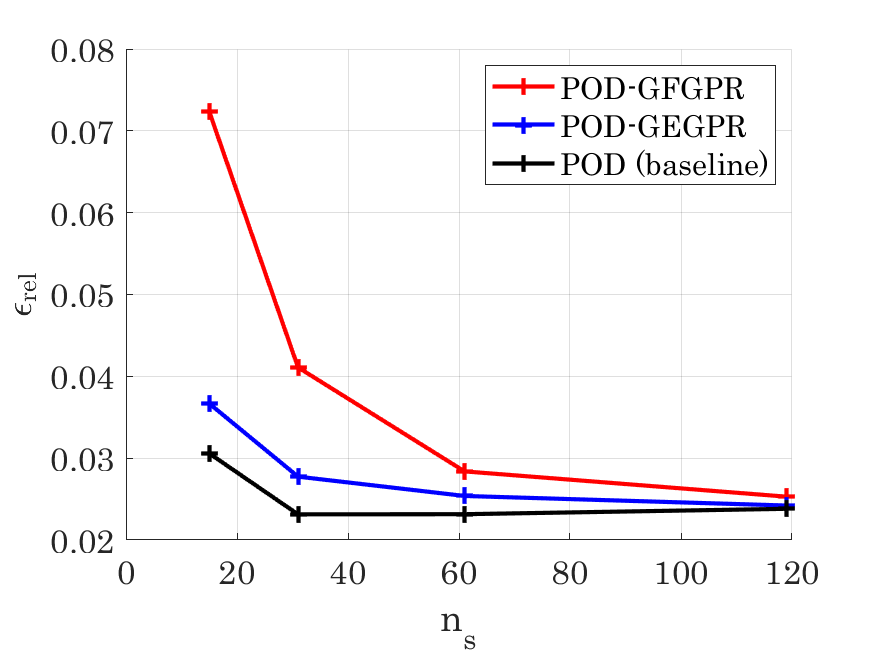}
	\caption{Average relative $L_2$ error of the magnetic flux density for gradient-free and gradient-enhanced GPR as a function of the number of samples in the training dataset, compared to the reconstruction error of POD.}
	\label{fig:pred_err}
\end{figure}

\begin{figure}[!t]
	\centering
	\includegraphics[width=0.9\linewidth]{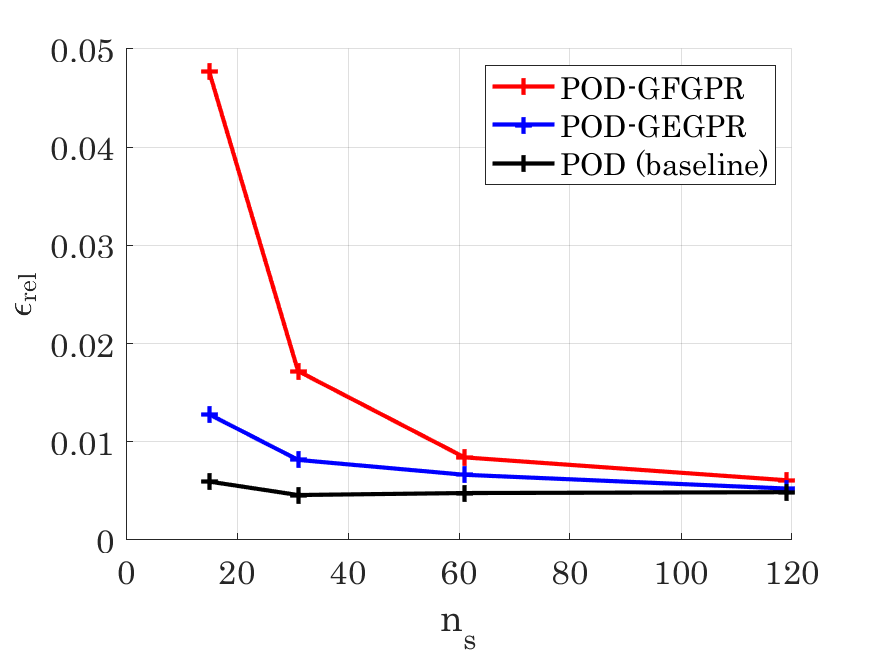}
	\caption{Average relative $L_2$ error of the magnetic flux density in the air region, for gradient-free and gradient-enhanced GPR as a function of the number of samples in the training dataset, compared to the reconstruction error of POD.}
	\label{fig:pred_err_air}
\end{figure}

\begin{figure*}[!t]
    \centering
    \begin{subfigure}[t]{0.3\textwidth}
        \includegraphics[width=\textwidth]{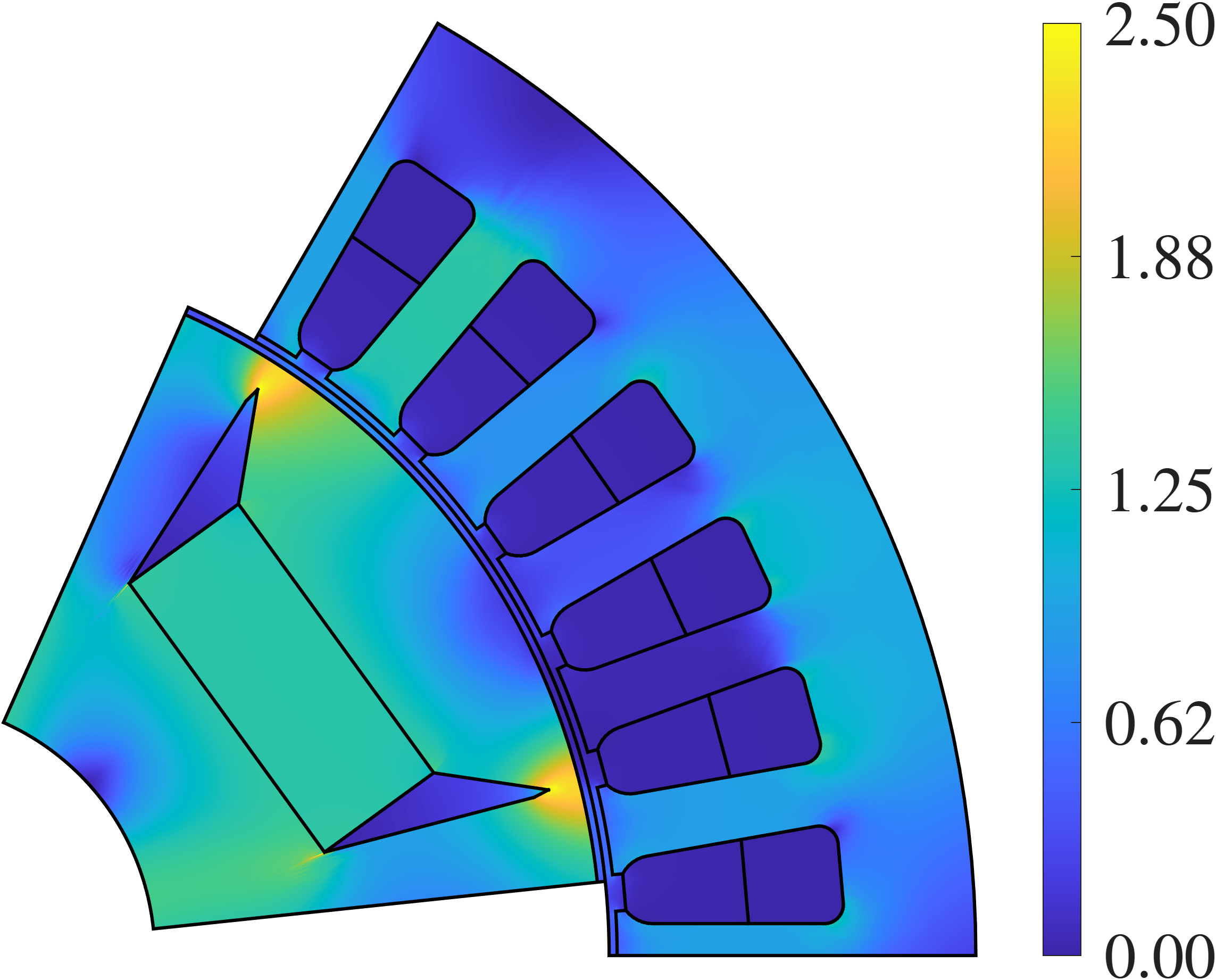}
        \caption{Reference magnetic flux density (T).}
    \end{subfigure}
    \hfill
    \begin{subfigure}[t]{0.3\textwidth}
        \includegraphics[width=\textwidth]{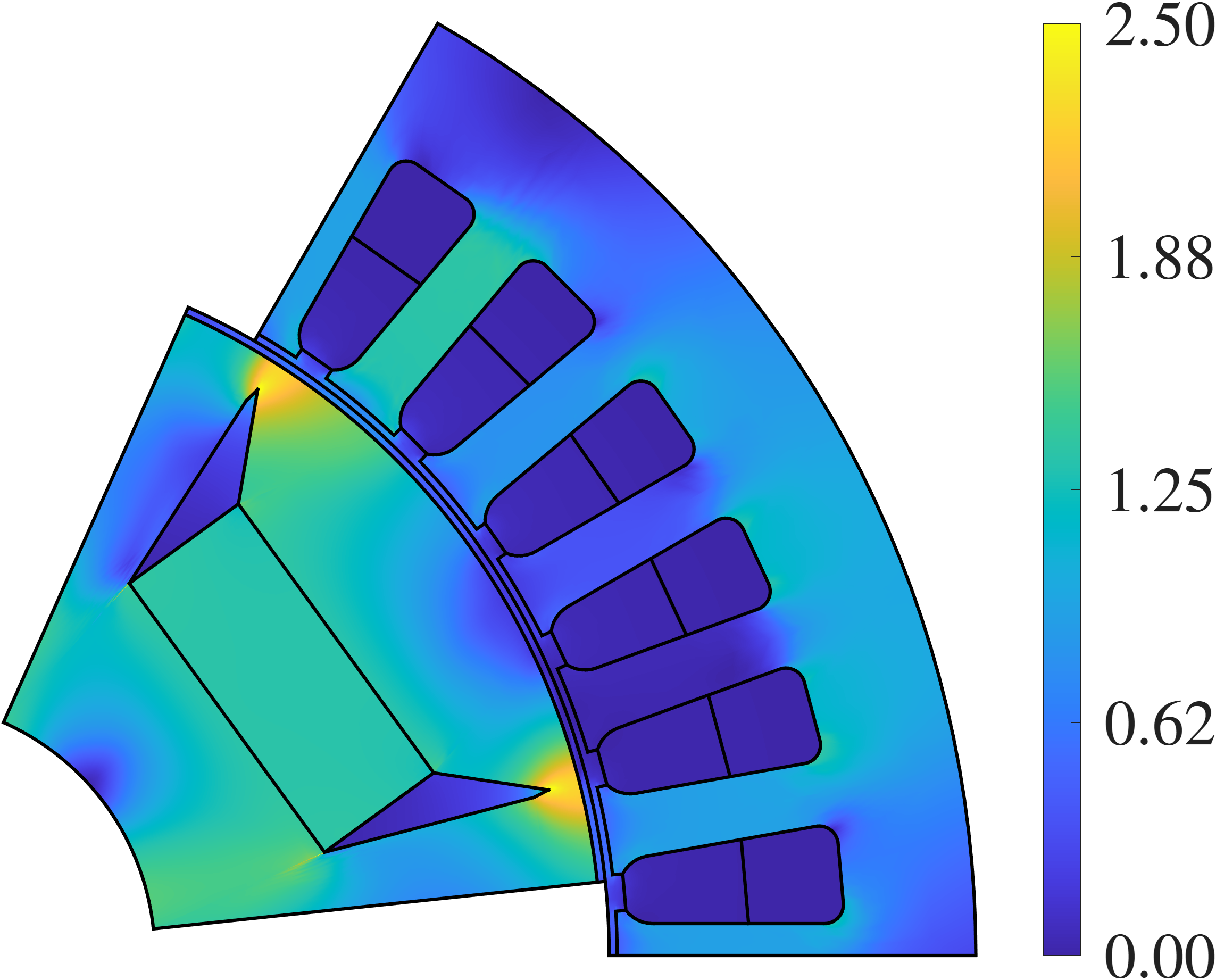}
        \caption{Predicted magnetic flux density (T).}
    \end{subfigure}
    \hfill
    \begin{subfigure}[t]{0.3\textwidth}
        \includegraphics[width=\textwidth]{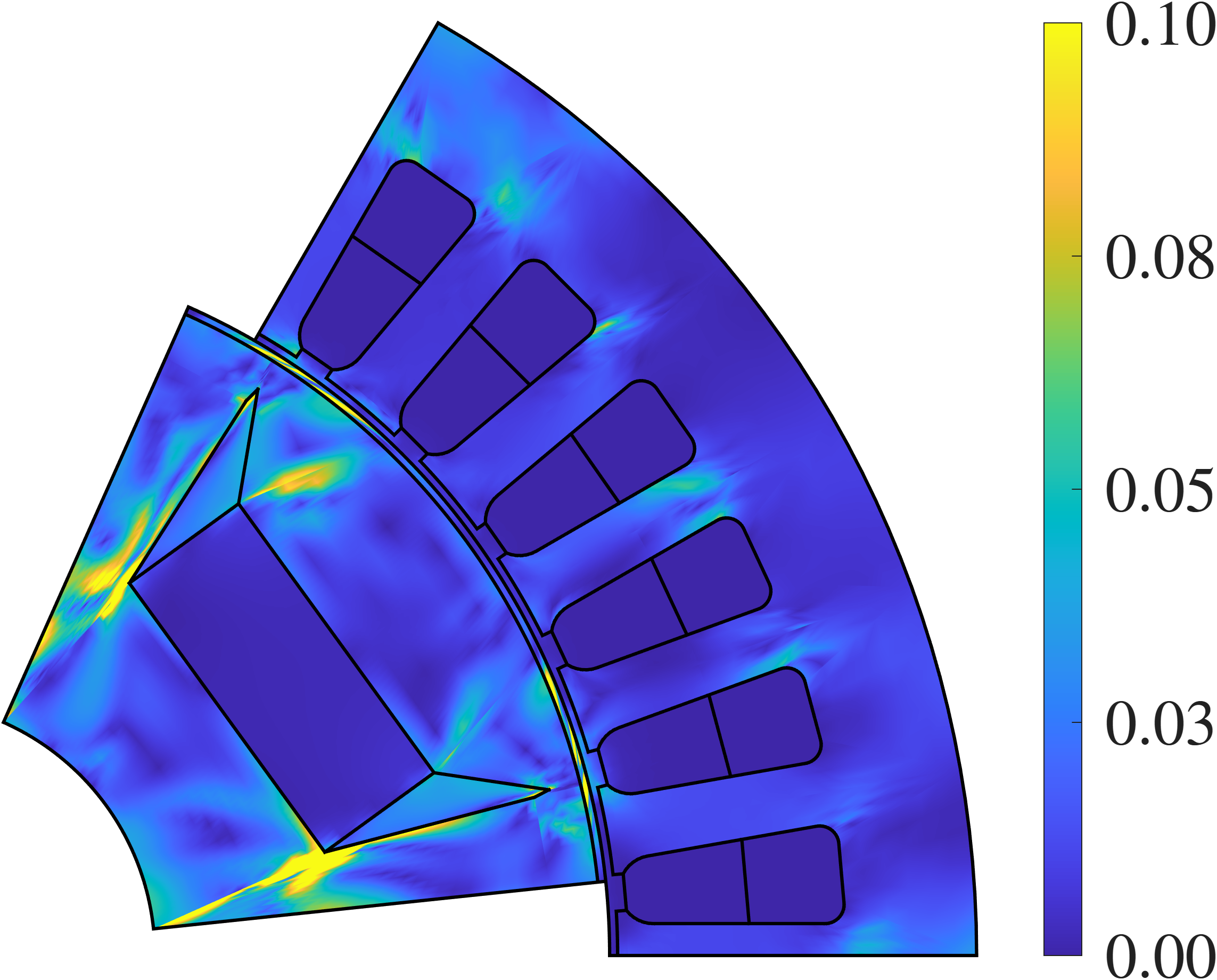}
        \caption{Absolute error in the magnetic flux density (T).}
    \end{subfigure}
    \caption{Comparison between true solution and the POD-GEGPR ($n_s= 31$) model's prediction for $\mathbf{p}$ = [MH = 8.47 mm, MW = 20.84 mm, MAG = 10.17 mm, $\theta_1$ = 21.68 $^\circ$] and $\alpha$ = 6 $^\circ$, corresponding to a relative error of 2.84\%.}
    \label{fig:fields}
\end{figure*}

Using the POD basis, the reduced coefficients $\bm{\xi}(\mathbf{p}^*)$ and their sensitivities are obtained for all parameter combinations in the training dataset. For each component of $\bm{\xi}$, two separate GPR models (GF and GE) are trained to map the parameter space to the corresponding reduced coefficient. Both input and output data are standardized, and the corresponding gradients are consistently rescaled to match the transformed variables. Models are then trained using the L-BFGS optimizer with 80 iterations and 3 restarts using GPyTorch~\cite{gardner2018gpytorch}. The two surrogate models are then compared by predicting the reduced coefficients $\tilde{\bm{\xi}}(\mathbf{p})$ starting from the parameter combinations in the test dataset. These predictions are subsequently used to reconstruct the full field solution $\tilde{\mathbf{u}}(\mathbf{p})$ via \eqref{eq:coeff_to_sol}, while the relative error of the magnetic flux density is computed through \eqref{eq:erel}, updating $\mathbf{A}$ for each parameter combination. As shown in \autoref{fig:pred_err} and \autoref{fig:pred_err_air}, increasing the number of training samples improves the accuracy of both the GF and GE models, with their errors approaching the POD reconstruction error, which constitutes a baseline for the model accuracy. This indicates that the reduced coefficients $\bm{\xi}$ are predicted with high accuracy and that the remaining error is primarily due to the POD truncation. The GE approach consistently outperforms the GF method, particularly in data-scarce regimes, motivating the use of gradient information for latent-space modeling. However, as reported in \autoref{tab:times}, this improved accuracy comes at the cost of increased training time due to the higher complexity of the model. A visual example of the error distribution of the POD-GEGPR model is presented \autoref{fig:fields} for a representative case with error close to the average.

\subsection{KPI prediction}
\begin{figure}[t]
    \centering
    \includegraphics[width=0.9\linewidth]{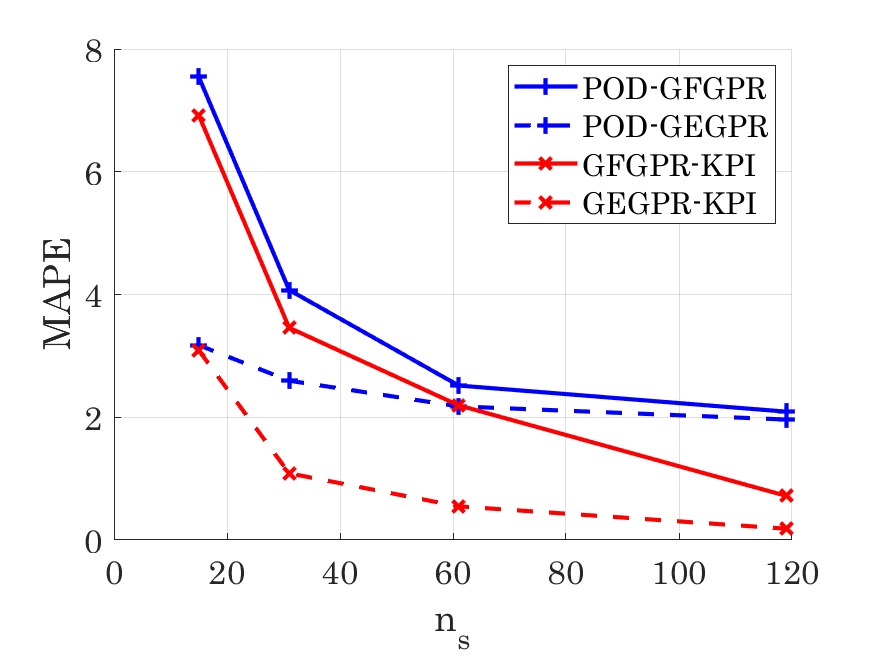}
    \caption{MAPE of the torque prediction from the gradient-free and gradient-enhanced POD-GPR models compared with GPR performed directly on the KPI (1\% corresponds to 0.01 Nm at 1 Nm).}
    \label{fig:torque_err}
\end{figure}

For the prediction of KPIs such as torque, two surrogate modeling strategies can be considered. The first consists of computing KPIs from the full-field surrogate model obtained in \autoref{sec:field_rec}. Alternatively, KPIs can be learned directly as functions of the input parameters using GPR, potentially leveraging the availability of torque sensitivities from the IGA model. For each rotor position, GF and GE GPR models are trained for torque prediction using the same standardization and training setup described in \autoref{sec:field_rec}. Their performance is assessed by comparing the Mean Absolute Percentage Error (MAPE) of the torque prediction, as shown in \autoref{fig:torque_err}. Results indicate that, while both methods yield accurate predictions, directly learning the parameter-to-torque map leads to superior accuracy, especially in data-rich regimes where the field reconstruction error constitutes the limiting factor of POD-GPR. In addition, this direct approach eliminates the need to store or construct the full IGA model, resulting in a more compact surrogate representation. Moreover, since only one output is required, as presented in \autoref{tab:times}, the training time is also decreased. Ultimately, this comparison is application-dependent, as it is influenced by the smoothness of the KPI and the reduced coefficients with respect to the parameter space, the truncation error, and the number of desired KPIs.

\section{Conclusion}
In this work, the differentiable nature of Isogeometric Analysis (IGA), enabling the efficient calculation of fields, Key Performance Indicators (KPIs), and their parametric sensitivities, is leveraged for efficient surrogate modeling of electric devices. Proper orthogonal decomposition is employed for dimensionality reduction, while gradient-enhanced Gaussian Process Regression (GPR)  learns a mapping between parameters and field coefficients to generate surrogate models for the magnetic flux density distribution. Moreover, GPR is also used for parametric surrogate modeling of KPIs. The approach is assessed on a parametric model of a permanent magnet synchronous motor, considering both data-scarce and data-rich regimes. 
Results demonstrate that, for both the magnetic flux density and the motor torque, including gradient information in the surrogate models significantly improves the model quality. While including gradient information increases the computational cost, it is justified in data-scarce regimes, where the available samples alone are insufficient to  capture the underlying behavior accurately. Moreover, the surrogate models reduce evaluation times from tens of seconds for the full IGA simulations to milliseconds, while maintaining high predictive accuracy. Given the short training times and the very fast prediction of new solutions, this approach represents a promising tool for parametric surrogate modeling of electric devices.

\section*{Acknowledgment}
This work was supported in part by the German Academic Exchange Service (DAAD) through the Program Research Grants in Germany, 2025, under Grant 57742125. This work was also supported by the Joint German Research Foundation (DFG) and the Austrian Science Fund (FWF) Collaborative Research Center CREATOR at TU Darmstadt, TU Graz, and JKU Linz under Project DFG: 492661287/TRR 361 and Grant
FWF: 10.55776/F90.
\ifCLASSOPTIONcaptionsoff
  \newpage
\fi

\input{references.tex}
\end{document}

%% file: figures/motor/PMSMgeometry.tikz
\tikzset{every picture/.style={line width=0.75pt}} %

\begin{tikzpicture}[x=0.75pt,y=0.75pt,yscale=-1,xscale=1]
\draw (330.53,226.06) node  {\includegraphics[width=380.3pt,height=337.08pt]{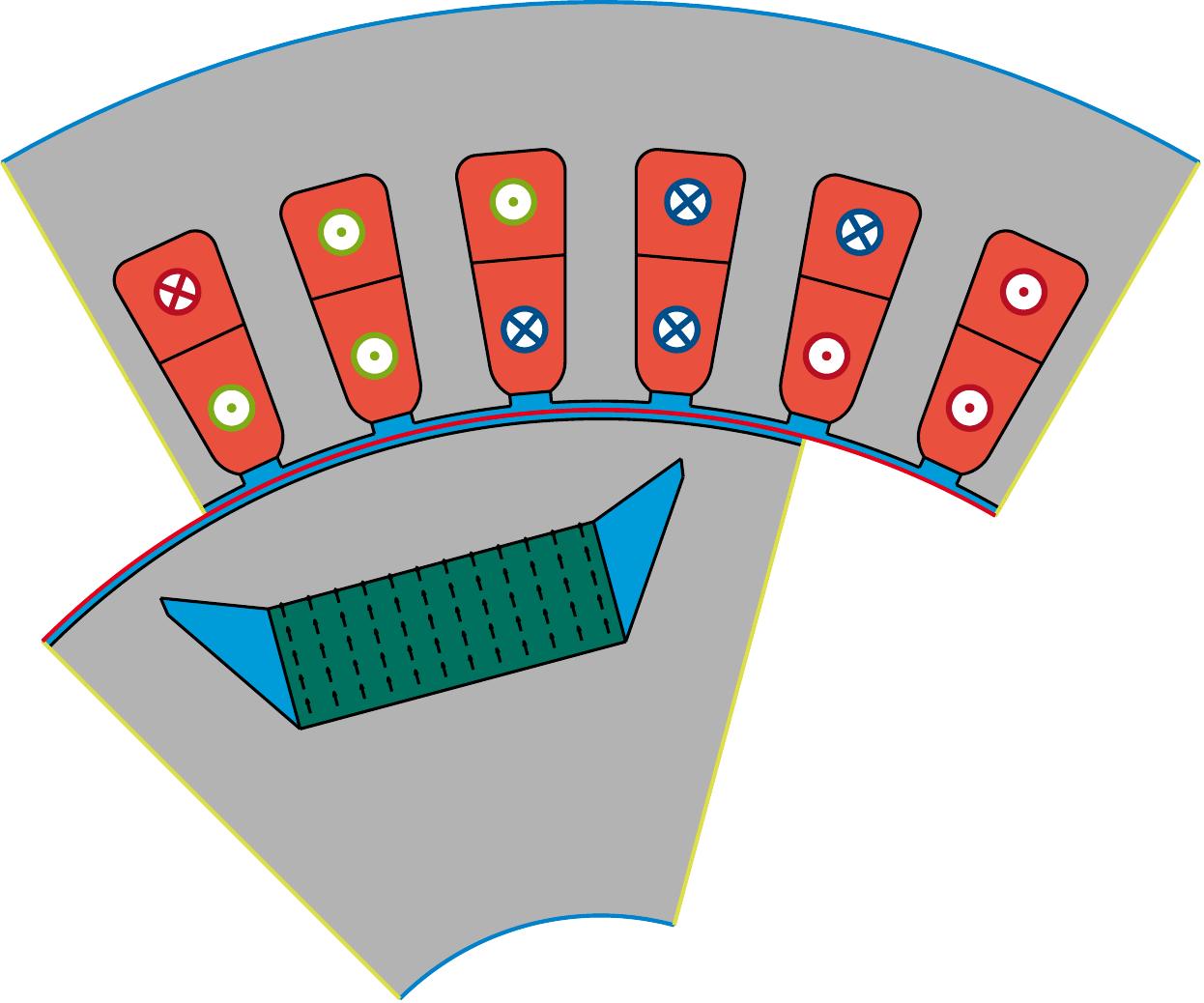}};
\draw    (208.22,333.96) -- (339.11,296.93) ;
\draw [shift={(342,296.11)}, rotate = 164.2] [fill={rgb, 255:red, 0; green, 0; blue, 0 }  ][line width=0.08]  [draw opacity=0] (8.93,-4.29) -- (0,0) -- (8.93,4.29) -- cycle    ;
\draw [shift={(205.33,334.78)}, rotate = 344.2] [fill={rgb, 255:red, 0; green, 0; blue, 0 }  ][line width=0.08]  [draw opacity=0] (8.93,-4.29) -- (0,0) -- (8.93,4.29) -- cycle    ;
\draw    (234.31,210.81) -- (240.63,233.69) -- (246.2,253.89) ;
\draw [shift={(247,256.78)}, rotate = 254.57] [fill={rgb, 255:red, 0; green, 0; blue, 0 }  ][line width=0.08]  [draw opacity=0] (8.93,-4.29) -- (0,0) -- (8.93,4.29) -- cycle    ;
\draw [shift={(233.52,207.92)}, rotate = 74.57] [fill={rgb, 255:red, 0; green, 0; blue, 0 }  ][line width=0.08]  [draw opacity=0] (8.93,-4.29) -- (0,0) -- (8.93,4.29) -- cycle    ;
\draw    (179.99,279.14) -- (193.2,326.55) ;
\draw [shift={(194,329.44)}, rotate = 254.44] [fill={rgb, 255:red, 0; green, 0; blue, 0 }  ][line width=0.08]  [draw opacity=0] (8.93,-4.29) -- (0,0) -- (8.93,4.29) -- cycle    ;
\draw [shift={(179.18,276.25)}, rotate = 74.44] [fill={rgb, 255:red, 0; green, 0; blue, 0 }  ][line width=0.08]  [draw opacity=0] (8.93,-4.29) -- (0,0) -- (8.93,4.29) -- cycle    ;
\draw    (248,205.73) .. controls (286.13,193.96) and (320.02,190.9) .. (360.51,194.86) ;
\draw [shift={(363,195.11)}, rotate = 185.95] [fill={rgb, 255:red, 0; green, 0; blue, 0 }  ][line width=0.08]  [draw opacity=0] (8.93,-4.29) -- (0,0) -- (8.93,4.29) -- cycle    ;
\draw [shift={(245,206.67)}, rotate = 342.33] [fill={rgb, 255:red, 0; green, 0; blue, 0 }  ][line width=0.08]  [draw opacity=0] (8.93,-4.29) -- (0,0) -- (8.93,4.29) -- cycle    ;
\draw  [dash pattern={on 0.84pt off 2.51pt}]  (241.67,201.78) -- (297.67,415.11) ;

\draw (281.33,316.27) node [anchor=north west][inner sep=0.75pt]   [align=left] {\LARGE MW};
\draw (202.85,239) node   [align=left] {\begin{minipage}[lt]{28.68pt}\setlength\topsep{0pt}
\LARGE MAG
\end{minipage}};
\draw (150.07,305.33) node   [align=left] {\begin{minipage}[lt]{21.28pt}\setlength\topsep{0pt}
\LARGE MH
\end{minipage}};
\draw (283.73,199.82) node [anchor=north west][inner sep=0.75pt]  [rotate=-354.67] [align=left] {\LARGE $\theta_1$};

\draw  [dash pattern={on 0.84pt off 2.51pt}]  (501,224.29) -- (425.37,350.64) ;
\draw    (411.55,233) .. controls (437.09,240.03) and (458.12,250.67) .. (477,263.29) ;
\draw [shift={(408.33,232.15)}, rotate = 14.53] [fill={rgb, 255:red, 0; green, 0; blue, 0 }  ][line width=0.08]  [draw opacity=0] (7.14,-3.43) -- (0,0) -- (7.14,3.43) -- cycle    ;

\draw (432,249.6) node [anchor=north west][inner sep=0.75pt]  [font=\LARGE]  {$\alpha $};

\draw  [fill={rgb, 255:red, 240; green, 240; blue, 240 }  ,fill opacity=1 ] (437,302) -- (582,302) -- (582,407) -- (437,407) -- cycle ;
\draw  [fill={rgb, 255:red, 233; green, 80; blue, 62 }  ,fill opacity=1 ] (443.57,332) -- (492,332) -- (492,352) -- (443.57,352) -- cycle ;
\draw  [fill={rgb, 255:red, 179; green, 179; blue, 179 }  ,fill opacity=1 ] (443.57,307) -- (492,307) -- (492,327) -- (443.57,327) -- cycle ;
\draw  [fill={rgb, 255:red, 0; green, 156; blue, 218 }  ,fill opacity=1 ] (443.57,357) -- (492,357) -- (492,377) -- (443.57,377) -- cycle ;
\draw  [fill={rgb, 255:red, 0; green, 113; blue, 94 }  ,fill opacity=1 ] (443.57,382) -- (492,382) -- (492,402) -- (443.57,402) -- cycle ;

\draw (507,309) node [anchor=north west][inner sep=0.75pt]   [align=left] {\Large Iron};
\draw (505,334) node [anchor=north west][inner sep=0.75pt]   [align=left] {\Large Copper};
\draw (506,359) node [anchor=north west][inner sep=0.75pt]   [align=left] {\Large Air};
\draw (504,384) node [anchor=north west][inner sep=0.75pt]   [align=left] {\Large Magnet};

\end{tikzpicture}